\documentclass[twocolumn,conference]{IEEEtran}
\usepackage{graphicx}
\usepackage{amsmath}
\usepackage{amssymb}
\usepackage{color}
\usepackage{paralist}
\usepackage{ifpdf}

\usepackage{caption} 

\usepackage{float}
\usepackage[utf8]{inputenc}
\usepackage{multirow}
\usepackage{rotating}
\usepackage{subfigure}
\usepackage{setspace}
\usepackage{amsmath}
\usepackage{moresize}

\usepackage{xcolor}

\usepackage[colorlinks,bookmarks=false]{hyperref}
\AtBeginDocument{%
  \hypersetup{
    citecolor=blue,
    linkcolor=blue,   
    urlcolor=blue}}

\usepackage{orcidlink}

\usepackage{url}
\usepackage{booktabs}
\usepackage{listings}
\usepackage{paralist}
\usepackage{wrapfig}
\usepackage{multirow}
\usepackage{ifpdf}
\usepackage{xspace}
\usepackage{keyval}
\usepackage{color}

\definecolor{listinggray}{gray}{0.95}
\definecolor{darkgray}{gray}{0.7}
\definecolor{commentgreen}{rgb}{0, 0.4, 0}
\definecolor{darkblue}{rgb}{0, 0, 0.4}
\definecolor{middleblue}{rgb}{0, 0, 0.7} 
\definecolor{darkred}{rgb}{0.4, 0, 0}
\definecolor{brown}{rgb}{0.5, 0.5, 0}

\usepackage[normalem]{ulem}
\makeatletter
\def\cyanuwave{\bgroup \markoverwith{\lower3.5\p@\hbox{\sixly \textcolor{cyan}{\char58}}}\ULon}
\def\reduwave{\bgroup \markoverwith{\lower3.5\p@\hbox{\sixly \textcolor{red}{\char58}}}\ULon}
\def\blueuwave{\bgroup \markoverwith{\lower3.5\p@\hbox{\sixly \textcolor{blue}{\char58}}}\ULon}
\font\sixly=lasy6 %
\makeatother

\newif\ifdraft
\ifdraft
\usepackage{xcolor}
\definecolor{ocolor}{rgb}{1,0,0.4}
\newcommand{\onote}[1]{ {\textcolor{ocolor} { (***Ole: #1) }}}
\newcommand{\terminology}[1]{ {\textcolor{red} {(Terminology used: \textbf{#1}) }}}

\newcommand{\jhanote}[1]{ {\textcolor{red} { ***shantenu: #1 }}}
\newcommand{\alnote}[1]{ {\textcolor{blue} { ***andreL: #1 }}}
\newcommand{\amnote}[1]{ {\textcolor{blue} { ***andreM: #1 }}}
\newcommand{\smnote}[1]{ {\textcolor{brown} { ***sharath: #1 }}}
\newcommand{\pmnote}[1]{ {\textcolor{brown} { ***Pradeep: #1 }}}
\newcommand{\msnote}[1]{ {\textcolor{cyan} { ***mark: #1 }}}
\newcommand{\mrnote}[1]{ {\textcolor{purple} { ***melissa: #1 }}}
\definecolor{orange}{rgb}{1,.5,0}
\newcommand{\aznote}[1]{ {\textcolor{orange} { ***ashley: #1 }}}
\definecolor{dandelion}{cmyk}{0,0.29,0.84,0}
\newcommand{\mtnote}[1]{ {\textcolor{dandelion} { ***matteo: #1 }}}
\newcommand{\note}[1]{}
\else
\newcommand{\onote}[1]{}
\newcommand{\terminology}[1]{}

\newcommand{\alnote}[1]{}
\newcommand{\amnote}[1]{}
\newcommand{\athotanote}[1]{}
\newcommand{\smnote}[1]{}
\newcommand{\pmnote}[1]{}
\newcommand{\jhanote}[1]{}
\newcommand{\msnote}[1]{}
\newcommand{\mrnote}[1]{}
\newcommand{\aznote}[1]{}
\newcommand{\mtnote}[1]{}
\newcommand{\note}[1]{}
\fi

\ifdraft
\usepackage{draftwatermark}
\fi

\newcommand{\pilotjob}{pilot-job\xspace}

\newcommand{\pilotabstraction}{pilot-abstraction\xspace}

\newcommand{\upp}{\vspace*{-0.5em}}

\lstdefinestyle{myListing}{
  frame=single,
  backgroundcolor=\color{listinggray},
  language=C,
  basicstyle=\ttfamily \footnotesize,
  breakautoindent=true,
  breaklines=true
  tabsize=2,
  captionpos=b,
  aboveskip=0em,
  belowskip=-2em,
}

\lstdefinestyle{myPythonListing}{
  frame=single,
  backgroundcolor=\color{listinggray},
  language=Python,
  basicstyle=\ttfamily \scriptsize,
  breakautoindent=true,
  breaklines=true
  tabsize=2,
  captionpos=b,
}

\ifpdf
\DeclareGraphicsExtensions{.pdf, .jpg, .tif}
\else
\DeclareGraphicsExtensions{.ps,  .eps, .jpg}
\fi

\usepackage{listings}
\usepackage{paralist}
\lstnewenvironment{code}[1][]%
{
\noindent
\minipage{1.0 \linewidth}
\vspace{0.5\baselineskip}
\lstset{
    language=Python,
    frame=single,
    captionpos=b,
    stringstyle=\ttfamily,
    basicstyle=\scriptsize\ttfamily,
    showstringspaces=false,#1}
}
{\endminipage}

\defaultleftmargin{1em}{}{}{}
\begin{document}

\title{Methods and Experiences for Developing Abstractions for Data-intensive, Scientific Applications}

\author{Andre Luckow$^{1,2,\orcidlink{0000-0002-1225-4062}}$  and Shantenu Jha$^{2,3}$\\
   \footnotesize{\emph{$^{1}$Ludwig-Maximilian University, Munich, Germany}}\\
   \footnotesize{\emph{$^{2}$RADICAL, ECE, Rutgers University, Piscataway, NJ 08854, USA}}\\
    \footnotesize{\emph{$^{3}$Brookhaven National Laboratory, Upton, NY, USA}\upp\upp\upp}
}

\date{}
\maketitle

\jhanote{for what?}\alnote{added the specific problem in here. I was not sure whether this problem should be motivated before referring to it} 
\jhanote{How is that
different from any other approach / method -- which would have design,
implementation and evaluation?}\alnote{added sentence}
\jhanote{and systems as well, I presume?}\alnote{fixed}
\jhanote{across scales?}\alnote{refined}
\jhanote{are we referring to DSR method? If not, what is process?}\alnote{refined}  

\begin{abstract}
Developing software for scientific applications that require the integration
of diverse types of computing, instruments, and data present challenges that
are distinct from commercial software. These applications require scale, and
the need to integrate various programming and computational models with
evolving and heterogeneous infrastructure. Pervasive and effective
abstractions for distributed infrastructures are thus critical; however, the
process of developing abstractions for scientific applications and
infrastructures is not well understood. While theory-based approaches for
system development are suited for well-defined, closed environments, they have
severe limitations for designing abstractions for scientific systems and
applications. The design science research (DSR) method provides the basis for
designing practical systems that can handle real-world complexities at all
levels. In contrast to theory-centric approaches, DSR emphasizes both
practical relevance and knowledge creation by building and rigorously
evaluating all artifacts. We show how DSR provides a well-defined framework
for developing abstractions and middleware systems for distributed systems.
Specifically, we address the critical problem of distributed resource
management on heterogeneous infrastructure over a dynamic range of scales, a
challenge that currently limits many scientific applications. We use the
\pilotabstraction, a widely used resource management abstraction for
high-performance, high throughput, big data, and streaming applications, as a
case study for evaluating the DSR activities. For this purpose, we analyze the
research process and artifacts produced during the design and evaluation of
the \pilotabstraction. We find DSR provides a concise framework for
iteratively designing and evaluating systems. Finally, we capture our
experiences and formulate different lessons learned.

\end{abstract}

\section{Introduction}

New scientific applications and discoveries are enabled by advanced data and
compute infrastructures, algorithms, and tools. Scientific progress
increasingly depends on driving forward the ability to support large-scale
computational and data demands of simulations in conjunction with data
processing, analytics, and machine
learning~\cite{fourthparadigm,midas_spidal_2019}. The complexity of
developing, deploying and scaling scientific applications arises from various
sources, in particular, the increasing heterogeneity that exists at all
levels, from hardware, infrastructure, middleware to
software~\cite{osti_1473756}.

Abstractions are crucial for scalable systems that hide internal complexities
and expose simple interfaces~\cite{10.5555/114768,shaw1984}. Designing useful
abstractions is challenging: hiding complexity does not automatically lead to
simple interfaces. The possible design space for abstractions is typically
vast, and there is no consensus on what constitutes effective abstractions.
Further, there are no accepted recipes to design and develop abstractions for
large-scale scientific distributed compute and data infrastructures.

The particular challenge addressed in this paper is the design of abstractions
for resource management on distributed  and heterogeneous compute and data
infrastructure. Currently, the scale and uptake of scientific, data-intensive
applications are hindered by a reliance on proprietary application and
systems-level resource management systems. These are often implemented using
rigid and ad-hoc approaches~\cite{pstar12}. A generalized abstraction that
helps overcome these limitations and enable scalable applications is needed.

\jhanote{A reference here would be useful /
imperative}\alnote{done}\jhanote{When motivating and framing a problem, self
references are not ideal or received well}

\jhanote{the last sentence does
not follow, IMO from the previous sentence.}\alnote{removed sentence}

While formal approaches maybe suitable for closed systems, they have
limitations for designing open, scientific distributed systems. Iivari
emphasizes that the \emph{``theory-with-practical-implications research
strategy has seriously failed to produce results that are of real interest in
practice~\cite{Iivari2007}''}.  DSR is an iterative approach to building,
evaluating, and refining software systems. While many research approaches
solely focus on theory and knowledge, DSR emphasizes practical relevance. It
realizes that complex systems need to be designed and evaluated in real-world
settings. By introducing a rigorous evaluation of the produced artifacts, DSR
provides generalizable knowledge that informs future design iterations, but
can also be transferred to other problems.

\jhanote{It is critical to explain why it is chosen, over alternative
approaches, which could be what? Note we do not have a related work section.}\alnote{done in paragraph above}

\jhanote{I would not make the following bold statement at this stage, before
discussing the methodology and evaluation criterion:}\alnote{moved} 

 \jhanote{The later has been defined / stated
already:}\alnote{removed} 

\jhanote{Design objective of what?}\alnote{fixed}

We propose the application of the design science research method
(DSR)~\cite{dsr} to the design of an abstraction and middleware for
distributed resource management.  Specifically, we apply the DSR
method to the process of designing  the
\pilotabstraction~\cite{pstar12}. Based on in-depth studies of different
applications, we define the design objective of the abstraction and system. Using the rigorous, iterative DSR process, we design, evaluate, and evolve
the abstraction from a compute-centric to an integrated abstraction for
managing compute and data resources and applications. In this paper, we
demonstrate the suitability of DSR for creating well-defined abstractions and
implement these in a real-world system. 

As part of DSR, we define different evaluation methods and criteria for
assessing the abstraction. For example, we investigate  the usability and
versatility of the abstraction in several case studies, e.\,g., in
ensemble-based simulations, MapReduce, and stream processing applications. We
use conceptual modeling to provide and validate our understanding of the
\pilotabstraction and the underlying mechanisms. Further, we study different
implementations of the abstraction concerning the performance and 
scalability using different types of applications, e.\,g., from the domains of 
genome sequencing and light source sciences.

This paper makes the following contributions: (i) it uses the DSR framework to
assemble a set of methods for the design and evaluation of abstractions, (ii)
it demonstrates the validity of DSR for designing and evaluation abstractions, such as the \pilotabstraction, \jhanote{I'm not sure what the ``studying'' the design of
the pilot-abstraction means?} \alnote{refined} (iii) it surveys publications related to the
\pilotabstraction and investigates the used methods for design and evaluation,
and (iv) it synthesizes the experiences gathered during this process in a set
of lessons learned. DSR was initially introduced in the domain of information
system research; we believe this is its first application to scientific
distributed computing.

This paper is structured as follows: %
We begin with an introduction of the methodology in Section~\ref{sec:dsr}, and 
continue with an investigation of scientific applications and their 
characteristics in Section~\ref{sec:applications}. The result is 
five application scenarios that the abstraction needs to address. We present the
\pilotabstraction in Section~\ref{sec:pstar}. In Section~\ref{sec:eval}, we 
discuss the methods used for evaluating the system. We discuss our
learnings and experiences of applying DSR in Section~\ref{sec:experiences}.

\section{Methodology}

\label{sec:dsr}

\alnote{Methodology: the study -- the description, the explanation and the justification of methods. Kaplan~\cite{Kaplan1964}. Methods: Techniques sufficiently general to be common to all sciences}

\begin{figure}[t]
  \centering
    \includegraphics[width=.49\textwidth]{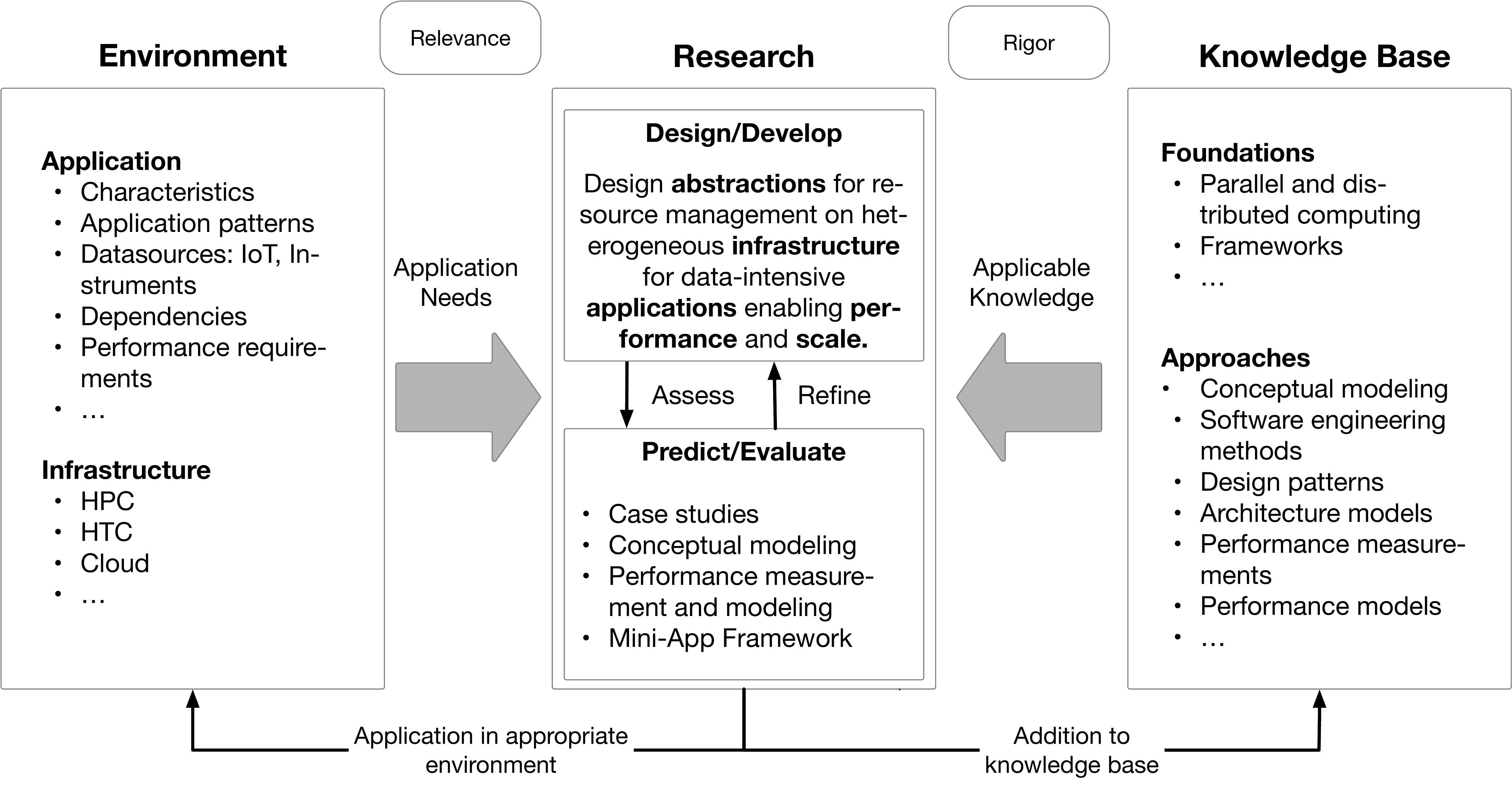}
  \caption{\textbf{Design Science Research Method (adapted from Hevner~\cite{dsr}):} To 
  address the complexity of the problem space, we follow an iterative research 
  approach of continuously building and evaluating abstractions.}
  \label{fig:figures_design_science_research}
\end{figure}

The objective of this section is to provide an introduction to \emph{design
science research (DSR)}~\cite{dsr}. 
DSR avoids the limitations of theory-based approaches, in particular, their
inability to capture complex, real-world systems. It emphasizes the iterative
creation, evaluation, and refinement of systems. The complexity of scientific
applications and infrastructure make DSR  suitable for designing
abstractions that enable applications to scale across heterogeneous
infrastructure. For this purpose, we customize  DSR  and apply it the
first time to the problem of abstraction development  (see Figure~\ref{fig:figures_design_science_research}).

The build-evaluate-refine cycle has two primary inputs: The environment provides
essential context for the problem, in particular, concerning application
requirements, characteristics, and infrastructure. The knowledge base defines,
in particular, the foundations and methodologies used the evaluation.
In the following, we give an overview of the different design science research
activities (adapted from Peffer~\cite{10.2753/MIS0742-1222240302}): (i) problem
identification, (ii) definition of objectives, (iii) design and implementation,
(iv) demonstration, and (v) evaluation.

\note{We develop a framework for capturing and analyzing
application characteristics, extracting commonalities and developing
abstractions and systems for as well as for evaluating these. This approach
provides the ability to capture high-level system aspects using conceptual
model. We evaluate the developed conceptual model and abstractions by implementing the system and a set of applications as
case study. Further, we utilize performance measurements and benchmarks methods
to evaluate the ability of the abstractions to capture important performance
trade-offs.}

\subsection{Problem Identification and Objectives}

Before starting the design process, an understanding of the problem
and design objectives is essential. Common methods for this activity are
literature reviews, expert interviews, focus groups, and
surveys~\cite{10.1007/978-3-642-29863-9_28}. In scientific application 
development, requirements frequently only emerge during the creation of the 
systems leading~\cite{oro17673}. Thus, iterative methods, such as DSR, are 
instrumental. The challenge addressed by this paper is the design and
development of effective abstractions that provide the right level of detail 
and support a variety of application scenarios and infrastructure while 
retaining ease-of-use.

\subsection{Design}

\subsubsection{Abstraction Design}
\label{sec:sysdesign}
\begin{figure}[t]
  \centering
   \includegraphics[width=.49\textwidth]{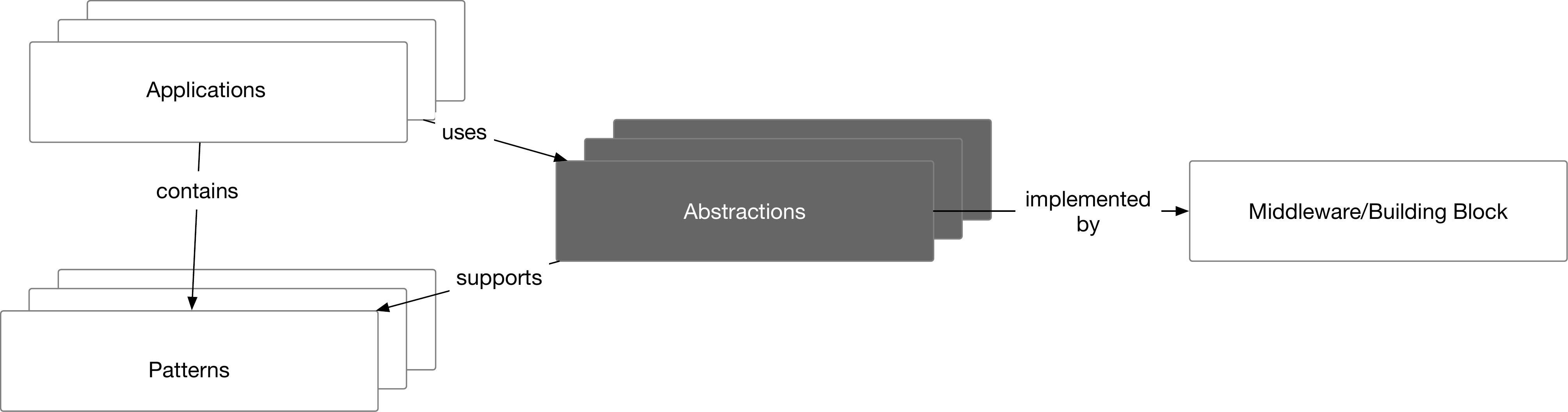}
  \caption{\textbf{Application Analysis (adapted from~\cite{doi:10.1002/cpe.2897}):}
  Patterns emerge by observing application characteristics and implementations. }
  \label{fig:figures_pattern_based_abstraction_development}
\end{figure}

Abstractions are a fundamental method of computer science, enabling reasoning
about a problem at the right level while allowing the underlying system to
implement a solution~\cite{10.5555/114768}. Shaw defines an abstraction as
\emph{``a simplified description or specification of a system that emphasizes
some of the system's details or properties while suppressing
others~\cite{shaw1984}.''}

To develop efficient abstractions, an understanding of applications and
infrastructure usage modes is instrumental. An important foundation for the
development of abstractions are patterns. Patterns are suitable solutions to
recurring problems in a particular context that can be applied multiple times
without doing it exactly the same
ways~\cite{Mattson:2004:PPP:1406956,doi:10.1002/cpe.2897}. Patterns can be
discovered by observing common problem decompositions (e.\,g., task and data
partitioning), communication, and coordination structures in applications. Jha
et\,al. utilize this process to study patterns and abstractions for distributed
applications~\cite{doi:10.1002/cpe.2897}. Mattson et
al.~\cite{Mattson:2004:PPP:1406956} investigate patterns for parallel and
distributed applications.
Figure~\ref{fig:figures_pattern_based_abstraction_development} illustrates the
relationship between applications, patterns, abstractions, and systems.
Discovered patterns serve as candidates for the development of abstractions
and their implementation in a middleware system.

An abstraction represents the external interface of the system. Thus, 
careful design is essential. The desired properties of an abstraction are
generality and simplicity~\cite{10.1145/3274770}. Generality 
refers to the ability of the abstraction to be broadly used. Simplicity is
reflected in multiple properties, e.\,g., ease-of-use, maintainability, and
extensibility~\cite{10.1145/1176617.1176622}.

The development of abstractions is a difficult task and requires the
identification of essential concepts, properties, and relationships. Conceptual
modeling enables abstract thinking and reasoning about a system and its
abstraction. Conceptual models represent and describe systems, e.\,g.,
applications, systems, and infrastructure. They can be used to formulate
concepts about the system, explaining how a system works. Mylopoulos defines a
conceptual model as `\emph{`a description of an aspect of the physical and
social world around us for the purposes of understanding and
communication~\cite{mylopoulos1992conceptual}.''} Johnson defines a conceptual
model as ``a high-level description of how a system is organized and
operates~\cite{Johnson:2002:CMB:503355.503366}.'' 
Conceptual models were introduced to computer science in 1984 by Brodie,
Mylopoulos, and Schmidt to overcome the increasing specialization of computer
science disciplines to describe high-level aspects and interactions
better~\cite{brodie1984conceptual}. Conceptual models are used in different
areas, e.\,g., for software architectures~\cite{Bruegge:2009:OSE:1795808}, and
programming languages~\cite{shaw1984}.

\subsubsection{System Design}

Software architecture and engineering research is the study of useful system
organization~\cite{919147}, i.\,e., the design of the
composition/decomposition of systems and subsystems and the communication
between these. Common objectives of the system design are
flexibility, maintainability, and re-usability~\cite{Bruegge:2009:OSE:1795808}. 
Patterns are an important aspect of designing software systems. Initially,
introduced in the domain of building architecture~\cite{alexander1977pattern},
were adopted to the domain of software architecture and engineering by
Beck/Cunningham~\cite{Beck1987}. 

A fundamental principle of system design is modularization and
decomposition. Modularization has many benefits, e.\,g., flexibility,
comprehensibility, and maintainability~\cite{Parnas:1972:CUD:361598.361623}.
Further, the development time can be reduced by the ability to distribute work
across different groups. 
According to Parnas, the most critical criteria for organizing
a system is information hiding~\cite{Parnas1971}, i.\,e., the ability to
carefully control the information exposed by a component using well-defined
external interfaces and hiding information that is likely to change.

A common type of modularization used by complex systems is a \emph{layered architecture}~\cite{Garlan:1994:ISA:865128}. The layered architecture model partitions the system in distinct hierarchical layers.
Each layer encapsulates a defined set of functions and provides services to the
layer above. This pattern is widely used in system-level software, such
as databases, operating systems, middleware, and distributed system. 

Similar to a layered architecture model is the hourglass model~\cite{10.1145/3274770}, which relies on a central bottleneck layer at the waist of the hourglass that connects a wide range of lower-level and higher-level services. Resource management is commonly described using  hourglass models~\cite{foster2003grid,toward-hpc-abds}.

\subsection{Evaluation}

A rigorous evaluation of all artifacts is a crucial part of the DSR process.
Sonnenberg/Brocke~\cite{10.1007/978-3-642-29863-9_28} propose four evaluation
activities. Eval 1 evaluates the problem
statements, using methods such as literature review and surveys. Eval 2
investigates the design specifications, e.\,g., using expert interviews, 
demonstrations, simulations, and benchmarking. Eval 3 is concerned with the
instantiation of an artifact, e.\,g., a prototype, using methods, such as
experimentation. Last, Eval 4 observes the artifact in the real world. We
utilize different evaluation methods for different activities, in
particular, case studies~\cite{10.2307/258557}, performance characterization,
and modeling~\cite{Jain91}. In the following, we particularly focus on methods for performance and scalability evaluations, i.\,e., Eval 3 and 4.

\subsubsection{Performance Characterization}

Performance measurements and characterizations are common methods for
describing a system in artificial and natural settings (Eval 3/4). 
Performance measurements can have different objectives: (i) workload and system
characterization, (ii) performance improvements, and (iii) to evaluate design alternatives~\cite{ferrari1978computer}. 
An essential component of a performance evaluation is the workload defined as the set of all inputs (programs and data) that a system
receives from its environment~\cite{ferrari1978computer}. A \emph{benchmark}
refers to a workload that is used to compare computer systems. A workload used
in performance evaluations should be representative; in the best case, it should
reflect an actual, real-world workload. \emph{Benchmarking} refers to the 
process of comparing two or more computer systems~\cite{Jain91}. 
A measure describes the performance of a system, e.\,g., the runtime or throughput of a
system. More complex metrics can capture cost/price or quality/runtime
trade-offs~\cite{Gray:1992:BHD:530588}.

Scientific applications are complex, unique, and not well-represented by 
standard benchmarks. 
The chosen metrics often do not provide a comprehensive view of the system, and thus, are not a proxy for
real-world performance~\cite{Gray:1992:BHD:530588}. Further, most benchmarks
neglect application-level quality metrics and focus mostly on runtime and
scaling performance. 
However, it is often challenging to obtain  real-world performance data that provide useful insights. 
For example, for data-intensive applications, there are, e.\,g., complex
infrastructure components, such as data source, broker, and processing
applications, that need to be carefully controlled. To account for that, often 
simplified, synthetic workloads are used to study performance (e.\,g., the 
Mini-App framework~\cite{DBLP:journals/corr/abs-1801-08648}). Similar techniques are commonly used for  
generating reproducible data and compute workloads,  
see~\cite{5388337,7004228,MERZKY2018329}.

\subsubsection{Performance Models}

Performance models~\cite{hoisie18} are a way of abstracting performance-related
insights into an analytical model. An analytical model is a precise
formulation of a model using mathematical logic, entities, and relations to
describe concepts~\cite{bordgida1984}. Analytical models are white-box  
and can quantify the relationship between the different concepts. Statistical models, in contrast, derive insights and predictions from
data~\cite{stat_model}.  The advantage of statistical models is that they do not
require domain knowledge and can model highly complex domains. However, they
are often black-box models, i.\,e., they are more difficult to interpret.

Many computer science domains use performance models, e.\,g., for  programming
languages, operating systems, database systems, system components (e.\,g.,
schedulers), as well as parallel and distributed systems. For example, database
systems utilize cost-based optimizer to generate an optimal query execution
plan~\cite{Selinger:1979:APS:582095.582099}. A well-known performance model for 
distributed applications is Amdahl's Law~\cite{Amdahl:1967:VSP:1465482.1465560}.

\alnote{Conceptual models are independent of the implementation, Formulative / 
Descriptive approaches, e.g. describing and proposing a specific abstraction, 
MapReduce, RDD Spark}

\alnote{Evaluation: show re-use through multiple applications}

\alnote{add mini app framework}

\alnote{Best practices for benchmarking and fair methods for summarizing results have
been extensively discussed (e.\,g., see Ousterhout~\cite{Ousterhout:2018:AMO:3234519.3213770} and Fleming et
al.~\cite{Fleming:1986:LSC:5666.5673}).}

\section{Application Scenarios, Characteristics and Requirements}
\label{sec:applications}

\note{Characterization vs. conceptual framework:
Characterization impact is local
Conceptual framework: impact is global
Characterization is biased by the sampling function
}

\note{SJ, 02/16/2020: 
\begin{itemize}
	\item Distinguish  scientific software  development challenge vs scientific 
	applications challenges.
	\item Distinguish  software (& programming) abstractions vs  abstractions 
	for conceptual / infrastructure / application / resource management ….
	\item Abstraction vs Abstract thinking: Identifying Pattern as candidates 
	for Abstraction is different from conceptual model which provides abstract 
	thinking. Currently Sec. II-B does a bit of both. If deliberate, maybe make 
	distinction explicit.
\end{itemize}
}

Understanding the problem domain is an essential step in the design process. In 
this section, we discuss important application characteristics and requirements.

\subsection{Application Characteristics}

The requirements of scientific applications are growing more diverse and
complex~\cite{DBLP:journals/corr/abs-1902-10810,NAP21886}. An increasing number
of instruments, such as light sources, telescopes, and  genome sequence machines,  generate vast volumes of data. Applications are becoming more sophisticated and increasingly require the combination of various processing types, e.\,g., simulation,
analytics, and machine learning. These processing types impose
different requirements on  abstractions, middleware, and
infrastructure.

While the integration of different processing types is challenging, it yields many
benefits, e.\,g., it has been demonstrated that machine learning-based
approximation techniques can improve simulations (e.\,g., by quickly identifying
regions of interests). Another example is the guidance of experiments using
machine learning, e.\,g., to find interesting events and regions, and to adapt
sampling accordingly~\cite{DBLP:journals/corr/abs-1902-10810}.

While data-intensive, scientific applications are highly diverse, they often share common computational and data characteristics.  Early studies, e.\,g., the Berkeley Dwarfs~\cite{dwarfs},
focused on the understanding of parallel algorithms based on their computation
and data movement patterns. Jha et\,al.~\cite{doi:10.1002/cpe.2897} study
distributed applications. In D3 science~\cite{doi:10.1002/cpe.4032}, we
conducted a survey consisting of 9 questions and a series of workshops to
understand the distributed and dynamic data aspects of 13 scientific
applications. The Big Data Ogres~\cite{bigdata-ogres,fox_bigdata_benchmarks}
introduce a multi-dimensional framework, so-called facets, which represent key
characteristics of big data applications and use them to define a set of Mini Apps based on a study of more than 50 use cases collected by
NIST~\cite{nist-uc}. 

\jhanote{Ref 48 and 60 are the same, i.e. same entry twice into different
references.}\alnote{fixed}

\begin{table*}[t]
	\centering \scriptsize
	\begin{tabular}{|p{1.5cm}|p{2.7cm}|p{2.9cm}|p{2.7cm}|p{2.7cm}|p{2.8cm}|}
	\hline
	&\textbf{Task-Parallel} &\textbf{Data-Parallel/MapReduce} &\textbf{Dataflow}	&\textbf{Iterative}	&\textbf{Streaming} \\ 
	\hline
	Description  &Focus on functional decomposition into tasks and control flow
				 &Decomposition based on data with minimal communication between tasks
				 &Multiple processing stages modeled with a directed 
				 acyclic graph  
				 &Multiple generations of tasks with sharing of data between the 
				 generations            
				 &Processing of unbounded data feeds in near-realtime
	\\ \hline
	Characteristics 	&Decomposition of a problem into a diverse set of  
						dependent and parallel tasks
						&Embarrassingly	parallel, loosely-coupled with minimal 
						communication. Details, such as communication and 
						synchronization hidden from the application
						&Multiple stages, loosely-coupled parallelism, global 
						communication for shuffle operation 
						&Loosely coupled parallelism with global communication 
						for updating machine learning model parameters
						&Data is processed in small batches often using 
						data-parallel algorithms. For many algorithms, a global 
						state needs to be maintained across batches of data
	\\ \hline
	Application Example &Molecular Dynamics~\cite{repex_ptrsa,doi:10.1021/ct4007037}, 
						Ensemble-Kalman Filter~\cite{El-Khamra:2009:DAD:1555301.1555304}, Scientific Gateways and Workflows~\cite{dare}
						&Map-Only analytics~\cite{hpc-abds}, Molecular Data 
						analysis Hausdorff Distance~\cite{Paraskevakos:2018:TAM:3225058.3225128}	
						&MapReduce for sequence alignment~\cite{Mantha:2012:PEF:2287016.2287020}, Molecular Data 
						analysis leaflet finder and RMSD~\cite{Paraskevakos:2018:TAM:3225058.3225128}	
						&Machine learning algorithms,
						K-Means~\cite{DBLP:journals/corr/JhaQLMF14}
						&Streaming for light source 
						data~\cite{DBLP:journals/corr/abs-1801-08648}%
						\\ \hline
	\end{tabular}
	\caption{\textbf{Data-Intensive Application Scenarios -- Characteristics and Patterns:}  Data-intensive applications are more complex than 
	compute-oriented applications and require the management of data, I/O and 
	compute resources. 
	\label{tab:applications} }
\end{table*}

Based on an investigation of 50+ applications and their characteristics
in~\cite{doi:10.1002/cpe.4032,bigdata-ogres,hpc-abds}, we derived five
applications scenarios: task-parallel, data-parallel, dataflow, iterative, and
streaming (see Table~\ref{tab:applications}). In the following, we discuss
important characteristics and patterns found in these application scenarios.

An important characteristic is the decomposition pattern: Task-level 
parallelism describes the execution of diverse
compute tasks on multiple compute resources. In contrast, data-parallelism
creates tasks by partitioning the data. Abstractions, such as 
MapReduce~\cite{Dean:2008:MSD:1327452.1327492}, enable
the data-parallel processing and aggregation of data using high-level
primitives. The runtime system then handles the implementation of the
parallelism, i.\,e., the partitioning of the data, the mapping of data to
tasks, the orchestration and synchronization of tasks and data
movements.

The dataflow model further generalizes the data-parallel model by supporting
applications comprising of multiple stages of processing.
The abstraction is based on directed acyclic graphs, where nodes
represent multiple stages of processing and the flow of data between
these stages. It was invented in the 1960s at MIT~\cite{sutherland1966line} and
later adapted to the domain of data-intensive computing (LGDF2~\cite{301911}, Dryad~\cite{dryad}) as a way to describe data
processing pipelines comprising of multiple stages, e.\,g., map, reduce,
shuffling. A stage can also be comprised of an external application (e.\,g., a 
simulation).

Iterative computation is a scenario applicable in particular to model training 
in machine learning applications.  An important requirement of these 
types of applications is the need to cache data to facilitate 
reading and processing data multiple times~\cite{Ekanayake:2010:TRI:1851476.1851593}.
In machine learning applications, this pattern is often found as many 
optimization techniques require multiple passes on the data to compute and 
update model parameters.

The last scenario is stream processing, defined as the ability to process
unbounded data feeds and provide near-realtime insights~\cite{streaming2015}.
The processing patterns for streaming are similar. However, the amount of data is typically smaller, as 
messages are processed in small batches. The management of
state between individual messages can be required. Stream processing is used to
analyze data streams from scientific experiments, e.\,g., light source
sciences~\cite{DBLP:journals/corr/abs-1801-08648}.

\jhanote{No surprises, I appreciate the classification of 5 types of data-intensive applications. However, what is unclear is how the following
paragraph about general requirements and the 4 specific (R1-R4) are generated
from or related to these 5 types?} \alnote{partitioned in two subsections}

\subsection{Application Requirements}

To support these scenarios, an efficient resource management abstraction and
middleware that can support highly diverse task-based workloads is required.
The heterogeneity of the tasks in these application scenarios is high, i.\,e.,
often tasks with diverse runtime, resource, and data requirements need to be
efficiently managed. For example, complex scenarios require the management of
both long-running tasks, e.\,g., simulation tasks, and short-running tasks,
e.\,g., data-parallel tasks arising from analytics applications. Further,
data-intensive applications can be highly unpredictable due to
data-dependencies and as a result very complex task graphs. The necessity to
respond to dynamic events demands support for task creation at runtime. The
requirements for the abstraction and middleware can be summarized as follows:
\begin{enumerate}[R1]
	\item \textbf{Abstractions:} Provide a higher-level abstraction that hides 
	the details of complex distributed infrastructure, but allows reasoning 
	about trade-offs. The abstractions should be \emph{simple} and 
	easy-to-use, 
	while supporting as many application scenarios as possible
	(\emph{applicable}). Further, it should be \emph{generalizable} to multiple 
	systems and implementations.
	
	\item \textbf{Middleware for Application-Level Resource Management:} 
	Provide the ability to manage highly diverse parallel 
	and dependent tasks and associated data on heterogeneous 
	infrastructure comprising of complex hardware and software stacks. The 
	system should support the \emph{interoperable} use of heterogeneous 
	infrastructures, in particular high performance computing (HPC), high 
	throughput computing (HTC) and cloud infrastructures,  and 
	should be \emph{extensible} to new frameworks and applications. 

	\item \textbf{Dynamism and Adaptivity:}  Ability to respond to changes in 
	the environment at runtime. Both middleware and abstraction need to support 
	this capability. 
	
	\item \textbf{Performance, Scalability, and Efficiency:} The system should 
	provide adequate performance, mainly high-throughput and low latencies, for 
	highly diverse task-based workloads. By doing so, the system supports  the  
	strong and weak \emph{scaling} of applications while ensuring efficient 
	resource usage.

\end{enumerate}

\alnote{very evolving field. lack of clarity is a first-order consideration. application characteristics is changing. different nomenclature for different application}

\section{Pilot-Abstraction: An Abstraction and Model for Distributed Resource Management}
\label{sec:pstar}

\alnote{ In the following, we attempt to answer
the following questions: What does it take to design software systems that
allow us to make faster and better advances in science? What abstractions are
needed to design systems at larger scale?
}

The need for tools and high-level abstractions to support application
development and the extreme heterogeneity of infrastructures has been widely
recognized~\cite{doi:10.1002/cpe.2897}. Resource management is a
fundamental challenge in distributed and parallel computing. The current state
is characterized by highly heterogeneous and fragmented systems, rigid point
solutions, and a lack of a unified model for expressing data and compute
tasks. The advent of data-intensive, machine learning, and streaming
applications complicated the state even further. Infrastructure is getting more
complicated by introducing new storage and memory tiers as well as accelerators.

\alnote{refer to application resource management needs: tasks, complex task dependencies, programming models and system and associated data}

Data-intensive applications exhibit complex characteristics and demand a highly
flexible abstraction for allocating resources and managing highly diverse
workloads of tasks. Balancing application characteristics and infrastructure
requires careful consideration of application-level and
infrastructure-level concerns. High-level abstractions are critical to retain
developer productivity and to scale applications. For example, the ability to
manage resources efficiently, taking into account application objectives is
important. By infusing application knowledge (e.\,g., about the data, compute
and I/O characteristics) into scheduling decisions, the runtime and scalability
can be significantly improved~\cite{prio-dnn}. Thus, data needs to be
integrated into these abstractions as first-class citizen.

\subsection{Pilot-Abstraction and Conceptual Model}
\label{sec:conceptual_model}

The pilot-abstraction~\cite{pstar12} is a unified abstraction for resource
management on heterogeneous infrastructure from high-performance computing,
high-throughput computing, big data, and cloud for distributed applications. In 
the following, we discuss the experience of developing and extending the 
abstraction and underlying middleware systems to support the described 
application scenarios. In this section, we focus on the DSR activities and artifacts related to the design and creation of the abstraction and middleware system.

We follow the iterative design approach of DSR closely aligning the abstraction
design to real application needs. The first system focused on the design of an
application-internal resource management framework for replica-exchange
simulations~\cite{repex_ptrsa}. The resource management
capabilities were evaluated using different application scenarios as case
studies and performance measurements on different
HPC infrastructures focusing on the internal resource management subsystem.
Based on the positive evaluation, we generalized the abstraction and created a re-usable, application-agnostic, standalone pilot-job system called
BigJob~\cite{saga_bigjob_condor_cloud}.

The observation of similar concepts for other infrastructures and
applications~\cite{glidein} motivated the development of the
pilot-abstraction~\cite{pstar12}, a general abstraction for the re-occurring
concept of utilizing a placeholder job as a container for a set of
computational tasks. The abstraction comprises of two main concepts, the pilot
that represents a placeholder for a specific set of resources and the
compute-unit, a self-contained task. The implementation of the pilot-job system
conceals details about the resource management systems of the different
infrastructures (e.\,g., HPC, HTC, and clouds). Thus, the user can focus on the
composition of tasks rather than dealing with infrastructure specific aspects.

The pilot-abstraction addresses the need to efficiently and flexibly manage
resources on application-level across distributed, heterogeneous
infrastructure. Pilot-jobs provide two key capabilities: (i) they support the
late binding of resources and workloads, and (ii) they provide a higher-level
abstraction for the specification of application workloads removing the need
the manage the execution of the workload manually. At the same time, they
provide critical capabilities to compose task- and data-parallel workloads
while providing optimal scalability and performance by managing task
granularities, data dependencies, and I/O via the abstraction. The design of  
the \pilotabstraction aims to offer a simple as possible and general interface 
to these capabilities.

Another artifact of the design process is a conceptual model for
understanding pilot-based systems. The P* model~\cite{pstar12} aims to
provide a common framework to understand the abstraction, as well as
commonly used \pilotjob systems. The P* model defines the high-level concepts
and mechanisms found in most pilot-systems. The model specification is done
using a functional description of the components and interactions, as well as various components and interaction diagrams. The characteristics and interactions of all concepts 
and analysis of different pilot-systems using the model is available 
in~\cite{pstar12}. 
Turilli et al.~\cite{turilli2017comprehensive} refined the framework.

\alnote{relate better to modeling}

While the pilot-job concept was developed for HPC and HTC, the need to
manage data in conjunction with pilots and tasks became apparent.
Pilot-Data~\cite{pilotdata} extends the pilot-abstraction and provides the
ability to manage storage and data, and couple these effectively with
computational tasks. With the emergence of big data frameworks, such as Hadoop, Spark, and Dask, the ability to couple HPC applications to specialized data processing engines become increasingly important, which lead to the development of Pilot-Hadoop~\cite{DBLP:journals/corr/LuckowMJ15,2016arXiv160200345L}. Further, extensions for in-memory processing~\cite{2016arXiv160200345L}, and
streaming~\cite{DBLP:journals/corr/abs-1801-08648} have been designed and
implemented.

\begin{figure}[t]
  \centering
    \includegraphics[width=.51\textwidth]{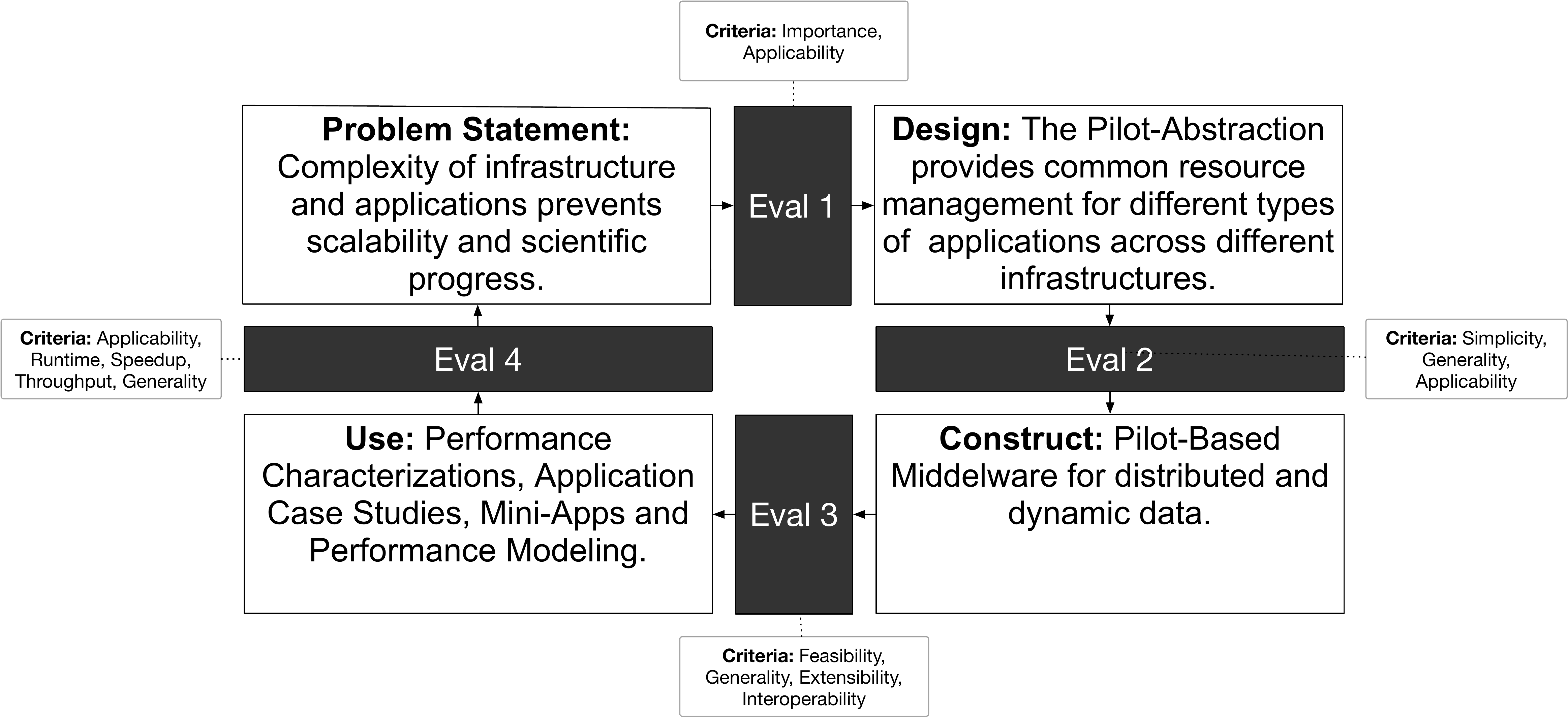}
  \caption{\textbf{DSR Evaluation Activities and Criteria (adapted from Sonnenberg~\cite{10.1007/978-3-642-29863-9_28}):} The incremental evaluation provides valuable input for refinement and valuable knowledge that can be transferred to other problems.
  \jhanote{I would redraw to use white space more effectively and thus be able to use larger font for text}\alnote{improved}}
  \label{fig:figures_dsr_eval}
\end{figure}

\subsection{Middleware: System Design and Architecture}

The objective of the system design phase is to create a system design and implementation that can support the desired abstractions. 
We applied methods and practices described earlier to achieve a flexible,
maintainable, and comprehensible architecture. The system architecture is based
on well-known design patterns~\cite{gamma1994design}, e.\,g., the adaptor
pattern for abstracting specific resource types, i.\,e., HPC, cloud,
and data infrastructures, such as Hadoop and Spark. For some of these
infrastructures, we utilize the SAGA~\cite{saga-x} as an access layer for local
resource management systems. The design artifacts of the architecture model are
created using block diagrams inspired by UML~\cite{uml}  to
visualize system layers, composition, and interactions. Examples of
architectural models artifacts can be found here:
Pilot-Job~\cite{saga_bigjob_condor_cloud}, Pilot-Data~\cite{pilotdata},
Pilot-Hadoop~\cite{DBLP:journals/corr/LuckowMJ15}, and
Pilot-Streaming~\cite{DBLP:journals/corr/abs-1801-08648}.

\section{Evaluation}
\label{sec:eval}

\alnote{speak to evaluation criteria: completeness, ease of use, effectiveness, efficiency, elegance, generality}
\jhanote{Should ``eval'' be uppercase?} \alnote{search/replace done}
\alnote{need to improve mapping of eval activities to table}

Evaluation is an essential part of the DSR process and ensures that the
designed system achieves the desired purpose. We explain the distinct types of
evaluation conducted on different artifacts produced throughout the Eval 1-4
activities proposed by Sonnenberg~\cite{10.1007/978-3-642-29863-9_28}.
Figure~\ref{fig:figures_dsr_eval} illustrates the four main activities:
problem statement, design, construction, and use. We discuss in-depth the used
methods and criteria used for evaluating the output of every stage.

Table~\ref{tab:eval} summarizes the evaluation methods used for the different DSR activities and the evolution of the \pilotabstraction. As proposed
in~\cite{10.1007/978-3-642-29863-9_28}, we evaluate the system interior,
i.\,e., the architecture, as well as the exterior, i.\,e., the usage of the
abstraction and system.

\begin{figure}[t]
  \centering
    \includegraphics[width=.4\textwidth]{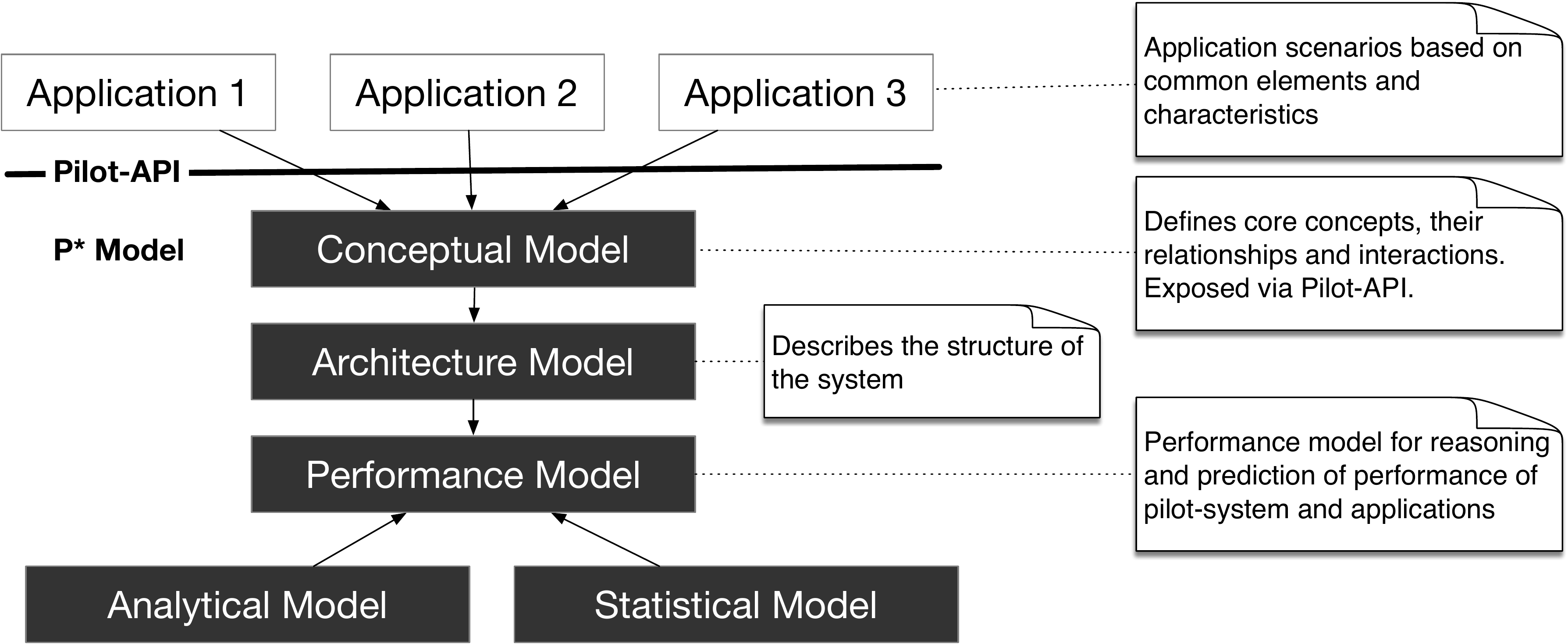}
  \caption{\textbf{Understanding the DSR Artifacts using Different Methods: } 
  Modeling techniques for characterization and evaluation of the
  \pilotabstraction and system. 
  \jhanote{I would redraw to use white space more effectively and thus be able
  to use larger font for text}\alnote{refined}}
  \label{fig:figures_Pstar_ConceptualModel}
   \label{fig:figures_model_hierarchy}
\end{figure}

Figure~\ref{fig:figures_model_hierarchy} summarizes the modeling methods used.
We use conceptual modeling to provide  high-level intuition and to allow
reasoning about inevitable trade-offs. Architecture models enable the
evaluation of the internal structure of the systems. Performance models are
used to describe the dynamic properties while using the abstraction and
system. Insights from the conducted evaluations inform the
abstraction design and to provide generalizable knowledge.  In the following,
we discuss the applied methods and criteria in detail.

\begin{table*}[t]
	\centering \scriptsize
	\begin{tabular}{|p{1.6cm}|p{3.8cm}|p{2.5cm}|p{2.5cm}|p{2.7cm}|p{2.7cm}|}
		\hline
					&\textbf{Pilot-Job~\cite{saga_bigjob_condor_cloud,pstar12}} &\textbf{Pilot-Data~\cite{pilotdata}} &\textbf{Pilot-Hadoop~\cite{DBLP:journals/corr/LuckowMJ15}} &\textbf{Pilot-Memory~\cite{2016arXiv160200345L}} &\textbf{Pilot-Streaming~\cite{DBLP:journals/corr/abs-1801-08648}}\\\hline		

Description &Management of computational tasks on heterogeneous infrastructure 
			&Management of data and compute tasks 
			&Management of Hadoop and Spark 
			&Management of in-memory runtimes for iterative tasks
			&Streaming data sources and processing\\\hline		

Infrastructure 
			&HPC, HTC, Cloud
			&HPC, Cloud, Hadoop/Yarn
			&HPC, Cloud, Hadoop/Yarn
			&HPC, Cloud, Hadoop/Yarn
			&HPC, Cloud, Serverless\\\hline

System Design\linebreak[4](Eval 2) &Conceptual model~\cite{pstar12}, 
				architecture model~\cite{saga_bigjob_condor_cloud}
		&Conceptual model~\cite{pstar12}, architecture model~\cite{pilotdata}
		&Architecture model~\cite{DBLP:journals/corr/LuckowMJ15} &Architecture model~\cite{2016arXiv160200345L} &Architecture model~\cite{DBLP:journals/corr/abs-1801-08648} \\\hline

Performance, Scalability and Efficiency \linebreak[4]  (Eval 3)
		&Pilot overhead, application and task runtimes, strong scaling, analytical model for 				replica-exchange simulations~\cite{async_repex11} 
		&Pilot overhead, application and task runtimes, strong scaling 
		&Runtime, strong scaling 
		&Runtime, strong scaling 
		&Throughput, latency, scalability, statistical performance model for throughput~\cite{luckow2019performance}\\\hline

Case Studies \linebreak[4]  (Eval 4) &Adaptive Replica Exchange~\cite{repex_ptrsa,async_repex11},
Ensemble Kalman Filter simulations~\cite{El-Khamra:2009:DAD:1555301.1555304},
HIV binding~\cite{doi:10.1021/ct4007037}, science portals~\cite{dare}, Pilot-MapReduce~\cite{Mantha:2012:PEF:2287016.2287020}
		&Genome Sequencing, K-Mean~\cite{pilotdata,DBLP:journals/corr/JhaQLMF14} &Wordcount, K-Means &K-Means &Light source data reconstruction, K-Means\\\hline	

	\end{tabular}
	\caption{\textbf{Evaluation:} Overview of Case Studies, Modeling Approaches and Performance Evaluation Methods Used. \label{tab:eval}}
\end{table*}

\subsection{Problem Identification and Design Evaluation (Eval 1/2)}

Eval 1 activity, i.\,e., the justification of the problem statement and
research gap, has been performed in the introduction and
Section~\ref{sec:applications}. The results of the literature and application
survey define the design objectives for the \pilotabstraction. The main
criteria applied for evaluation of the problem was the importance and
applicability of the design idea to a broad set of applications.

The design of the \pilotabstraction and middleware system is evaluated
according to three main criteria: simplicity, generality, and applicability. An important artifact of the design phase is the P* conceptual model. The model
defines the elements, characteristics, and interactions. The objective of P* is
to provide a minimal but complete model that provides an intuition of the
system. A metric for the simplicity of the model is the number of elements of the model, which is very low with four main concepts.
The design of the \pilotabstraction reduces the amount of code necessary
significantly while providing interoperability across different infrastructures. Further, we demonstrate the model's generality by comparing and
mapping different implementations of the \pilotabstraction~\cite{pstar12}.

\subsection{System Implementation (Eval 3)}
\label{sec:miniapp}

The Eval 3 activity evaluates the pilot-abstraction in artificial settings. The
developed conceptual models provide an important basis for the construction of
the system and the performance evaluation by offering essential information
about the structure and expected behavior of the system.

The prototype implementation of the pilot-system is evaluated using an 
architecture model comprising of several component and interaction diagrams. 
The main criteria are feasibility, extensibility,  interoperability. 

The feasibility and generality of the abstraction is shown in various
prototype and production
implementations~\cite{saga_bigjob_condor_cloud,10.1007/978-3-030-10632-4_4}.
Various extensions, e.\,g., for data management, in-memory processing, and in
support of new infrastructures, such as cloud and serverless, demonstrate the extensibility of the system. The
implementation maps the \pilotabstraction to the different infrastructures
enabling interoperability. We verified the interoperability by various
experiments with a broad set of distributed HPC and data-intensive applications.

\subsection{Performance and Case Studies (Eval 4)}
\label{sec:eval_results}

An important objective of the pilot-abstraction is to overcome barriers to
scaling. Thus, performance and scalability are essential evaluation criteria as
both are instrumental for the many scientific applications. We use three
approaches: (i) performance characterization of the pilot-system and several
applications, and (ii) analytical performance modeling and (iii) statistical
performance modeling for selected use cases.

As benchmarks do not correctly reflect the requirements of scientific
applications, we rely on custom experiments for evaluations. A challenge for
performance characterizations and modeling is the experimental design and data
collection. The experimental design is the process of determining the factors,
factor levels, and combinations of these for an experiment to understand the
effect of each factor while minimizing the number of
experiments~\cite{Jain91,hauck2014automated}. A good experimental design is
essential to capture essential characteristics while minimizing data collection
efforts.

\alnote{Add Mini-Apps
\begin{itemize}
	\item focus on applications  and abstractions understanding: what performance trade-offs should an abstraction be able to capture
	\item comparison to other systems not in scope
	\item 
	\item Fundamental characteristics that can be captured by Mini-Apps
	\item measures: runtime and scalability, in case of streaming throughput
	\item Judging the quality of the output is highly application dependent. Thus, we propose a mini-app framework to generate synthetic applications that can capture important characteristics of the main applications.
	\item Costs of measuring a real-world application are usually prohibitive
	\item 
	\item The objective for the Mini-App framework were derived from Gray: 
	relevant, portable, scalable and simple~\cite{Gray:1992:BHD:530588}.
	\item standard benchmarks often do a bad job of predicting a system's 
	performance in real-world conditions.
	\item Benchmark engineering~\cite{diss_mods_00016245}, p99, design approach 
	for correctly measuring data
	\item Synthetic data allows that everyone can generate data and replicate 
	benchmarks
\end{itemize}}

We propose the Mini App framework~\cite{DBLP:journals/corr/abs-1801-08648} to address these challenges and to automate and accelerate the build-assess-refine cycle. The Mini App framework helps to evaluate
abstractions, middleware, and infrastructure in real-world conditions.  Further, the data collected can serve as a basis for statistical models
and predictions. It was designed to support an excellent experimental design following
best practices defined by Gray~\cite{Gray:1992:BHD:530588} and
Waller~\cite{diss_mods_00016245}:
\begin{inparaenum}[(i)]
	\item \textbf{Simplicity:} Easy-to-use and setup via high-level APIs and 
	configurations.
	\item \textbf{Relevance:} It gives the developer full control of the 
	application workload and metrics necessary for the application scenario.
	\item \textbf{Scalability:} Support for distributed resources and datasets  at 
	various scale levels and data rates.
	\item \textbf{Portability:} Infrastructure and application-agnostic by design. 
	Different types of infrastructure supported via pilot-abstraction.
	\item \textbf{Reproducibility:} It provides comprehensive automation of 
	performance experiments ensuring repeatability and reproducibility.
\end{inparaenum}

Another important aspect of DSR is the ability to derive knowledge and insights. We use different modeling approaches to generalize  abstractions, systems, and applications. For example, we provide analytical
models for the performance of the application and
pilot-systems~\cite{async_repex11,pilotdata}. These models capture the
significant components of the runtime and allow users to understand the impact
of input data volume and parallelism on the runtime. Further, it enables the
assessment of the system overheads and their ratio to the overall
runtime of the application. Further, we use statistical modeling, e.\,g.,
for the prediction of the throughput of streaming systems for different
infrastructure configurations~\cite{luckow2019performance}.

Further, we evaluate the applicability of the abstraction in a natural setting,
e.\,g., in various
applications~\cite{El-Khamra:2009:DAD:1555301.1555304,10.1145/2484762.2484819},
and frameworks~\cite{Mantha:2012:PEF:2287016.2287020}. In these investigations,
we assess whether the \pilotabstraction meets the defined requirements 
concerning its capabilities, simplicity, and the feasibility to implement,
deploy, and execute applications. In particular, we focus on the resource
management requirements, such as the ability to adapt to changing resource
needs, while providing adequate performance and scalability. The abstraction
proved useful to capture the critical parameters necessary to express task and
data decompositions and the associated performance trade-offs. In various case
studies, we demonstrated that the abstraction allows a suitable control of the
compute  and data movements.

\note{The statistical model will be used to quantify the relationship between
this terms. The hybrid approach provides many advantages: the analytical model
is particularly useful for evaluation of the system design, while the
statistical model can quantify the relations between components and capture
more complex system behavior.}

\section{Discussion}
\label{sec:experiences}

Abstractions are vital for handling complexity and
building systems at an unprecedented scale. We present a balanced approach
using the design science research method to design and evaluate the
\pilotabstraction, an abstraction for enabling resource management across
heterogeneous, distributed resources. By iteratively addressing real-world
application and system challenges using DSR as a methodological framework, we
were able to develop and refine the \pilotabstraction. The incremental evaluation of the artifacts of the DSR process provides valuable input for future iterations and generalizable knowledge for similar problems. 

Using DSR, we designed and developed the \pilotabstraction, and evaluated it against the defined requirements:
\begin{enumerate}[R1]
	\item \textbf{Abstractions:} The Pilot-Abstraction's capabilities and
simplicity have been evaluated and validated in several application scenarios,
e.\,g., ensemble simulations, data-intensive applications, and streaming.
Further, the extensive usage of the Pilot-Abstraction for higher-level
building blocks, e.\,g., a workflow framework~\cite{turilli2019middleware}, an
ensemble simulations management framework, and a MapReduce
framework~\cite{Mantha:2012:PEF:2287016.2287020}, demonstrates its viability
and usefulness.

	\item \textbf{Middleware for Application-Level Resource Management:}
  The pilot-system provides interoperable use  of HPC, cloud, and data
  infrastructures.  In~\cite{cloud_book}, we explore the interoperable use of
  HPC, HTC, \jhanote{HTC not defined. Should we?}\alnote{good catch. Somehow the paragraph about the infrastructure types got lost in the shortening process} and clouds.
  \jhanote{Is reference to pilotdata properly formed for you?}\alnote{yes}
  In~\cite{pilotdata}, we use and characterize the use of Pilot-Data on HPC
  and HTC resources. The system is extensible to new infrastructures, such as
  Hadoop~\cite{DBLP:journals/corr/LuckowMJ15},
  streaming~\cite{DBLP:journals/corr/abs-1801-08648}, and
  serverless~\cite{luckow2019performance}.
	
	\item \textbf{Dynamism and Adaptivity:}  An important capability of the 
	\pilotabstraction is the ability to respond to changes in the	environment 
	at runtime. In~\cite{saga_bigjob_condor_cloud}, we explore the usage of 
	additional cloud 
	resources at runtime to meet application demands. 
	In~\cite{luckow2019performance}, we demonstrated a 
	model for throughput prediction to determine the optimal set of resources 
	for a given workload.
	
	\item \textbf{Performance, Scalability, and Efficiency:} We demonstrated in 
	various studies that the \pilotabstraction enables the creation of 
	scalable applications by given fine-grained control on data/task 
	composition 
	while hiding the details~\cite{saga_bigjob_condor_cloud, pilotdata,Paraskevakos:2018:TAM:3225058.3225128, luckow2019performance}. 	
\end{enumerate}

\begin{figure}[t]
  \centering
    \includegraphics[width=.33\textwidth]{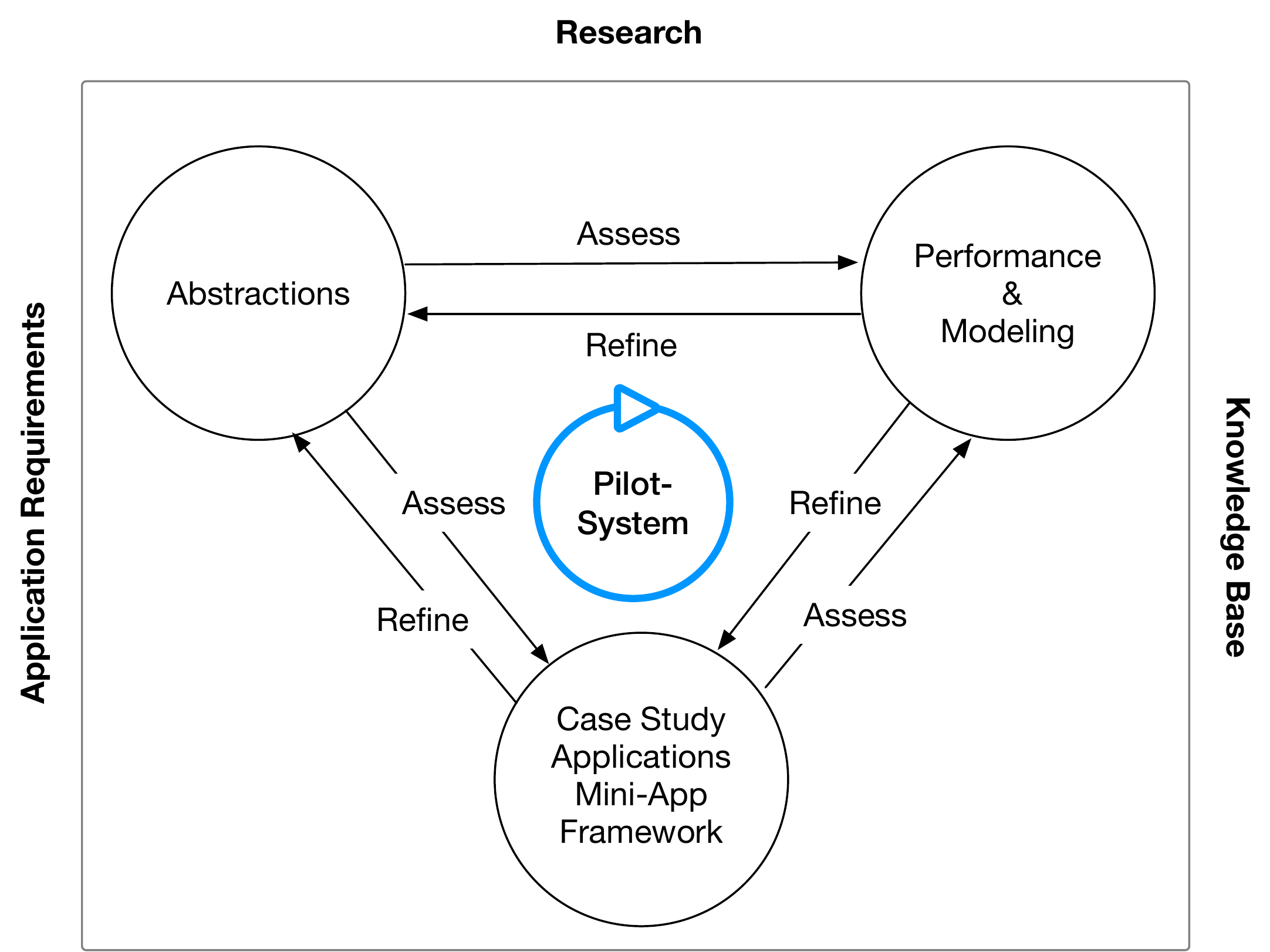}
  \caption{\textbf{Iterative Research Approach:} Using an iterative feedback 
  loop of abstraction design and evaluation using real-world and 
  synthetic applications to refine design and system.}
  \label{fig:figures_method_overview}
\end{figure}

In the following, we describe and synthesize our experiences of the
development of the \pilotabstraction in a set of lessons learned to inform
the design process of future systems.

\textbf{Iteration:} The iterative design and evaluation process of DSR is
instrumental in creating appropriate abstractions and middleware systems.
Building real systems and applications is instrumental in discovering new usage modes and further requirements. Implementing smaller working
systems is instrumental before scaling to more extensive resources and further
applications. Specifically, we iteratively grew the pilot-job system from
supporting coarse-grained ensembles of simulation tasks on single
infrastructures to support for high-volume, fine-grained data-parallel tasks,
and streaming.

\textbf{Automation:} Collecting data on the design is an instrumental part of
the process. Automating experiments for performance characterizations and
measurements is important to enable the exploration of larger parameter spaces
and to ensure reproducibility. We developed the Mini Apps framework to
formalize and automate the experiments and data collection.
Figure~\ref{fig:figures_method_overview} illustrates the feedback loop used for
the design of \pilotabstraction and the implementation in the pilot-system. By
using continuous evaluations, partially automated with the Mini App framework,
valuable inputs for the abstraction and experimental design and modeling process
are generated.

\textbf{Abstraction Design:} The design process is complex and requires the
careful trade-off of capabilities, simplicity, and generality. The more
application-specific knowledge can be induced via abstractions into middleware
systems, the better the decision the system can make, e.\,g., concerning
scheduling. However, the more application-specific the abstraction, the less
general is its utility. Balancing simplicity, generality, and capability is
challenging and requires a careful evaluation of the abstraction in 
different applications and settings.

\textbf{Compute and Data:} Managing heterogeneous compute tasks at scale is 
challenging by itself. The addition of data complicates the problem 
significantly. There is a significant amount of heterogeneity and dynamism in the way data can be stored, transferred, and used. Typically, a great 
extent of the data lifecycle is external to the applications. We address these 
challenges, particularly by focusing on defined application scenarios (see 
Table~\ref{tab:applications}) and by supporting and optimizing for important 
patterns, e.\,g., MapReduce.

\textbf{Optimize Application Algorithms:} A universal abstraction and system for
resource management can help to scale applications by simplifying and
standardizing the process of resource and task management. In many cases, an
improvement of algorithms can lead to even more significant improvements
compared to scaling out a non-optimal algorithm to more resources (see
e.\,g.~\cite{Paraskevakos:2018:TAM:3225058.3225128}).

\textbf{Limitations of Abstractions:} In many cases, systems are not limited by
conceptual abstraction, but by the implementation of the system and
infrastructure. Further, abstractions can exhibit undesirable behaviors. Leaky abstractions describe the phenomena that abstractions
frequently fail in real-world settings exposing complexities from underlying systems that it meant to abstract~\cite{leaky}.

\textbf{Re-Use and Interoperability:} A well-designed abstraction is a
minimal requirement for developing robust and scalable software
systems. By abstracting commonalities between systems, interoperability can be
achieved. However, significant investments into the stability and robustness of
the system are required to support real-world applications.

\note{The main objectives for this model are to support the following activities:
\begin{compactitem}
	\item \emph{System Understanding:} Support reasoning about  
	architecture trade-off.
	\item \emph{Resource Configurations:} Find a suitable resource 
	configuration for execution of a certain workload.
	\item \emph{Runtime Prediction:} predict the runtime of am application 
	based on set of parameters, e.\,g., input data size, computational 
	characteristics, scalability.
	\item \emph{Adaptive and Dynamic Execution:} Utilization of performance 
	model to dynamically	provision resources to meet processing SLAs.
\end{compactitem}
}

\bibliographystyle{unsrt}
\bibliography{experimental_design,radical_publications,local,bigdata,streaming,benchmark,pilotjob,saga-related}

\begin{thebibliography}{10}

\bibitem{fourthparadigm}
Anthony J.~G. Hey, Stewart Tansley, and Kristin~M. Tolle.
\newblock {\em The Fourth Paradigm: Data-Intensive Scientific Discovery}.
\newblock Microsoft, 2009.

\bibitem{midas_spidal_2019}
Geoffrey Fox, Judy Qiu, David Crandall, Gregor~Von Laszewski, Oliver Beckstein,
  John Paden, Ioannis Paraskevakos, Shantenu Jha, Fusheng Wang, Madhav Marathe,
  Anil Vullikanti, and Thomas Cheatham.
\newblock Contributions to high-performance big data computing.
\newblock In L.~Grandinetti, G.R. Joubert, K.~Michielsen, S.L. Mirtaheri,
  M.~Taufer, and R~Yokota, editors, {\em Future Trends of HPC in a Disruptive
  Scenario}. IOS Press Volume 34 of Advances in Parallel Computing, 2019.

\bibitem{osti_1473756}
Jeffrey~S. Vetter, Ron Brightwell, Maya Gokhale, ..., and Jeremiah Wilke.
\newblock Extreme heterogeneity 2018 - productive computational science in the
  era of extreme heterogeneity: Report for doe ascr workshop on extreme
  heterogeneity.

\bibitem{10.5555/114768}
Alfred~V. Aho and Jeffrey~D. Ullman.
\newblock {\em Foundations of Computer Science}.
\newblock Computer Science Press, Inc., USA, 1992.

\bibitem{shaw1984}
Mary Shaw.
\newblock {\em On Conceptual Modelling: Perspectives from Artificial
  Intelligence, Databases, and Programming Languages}, chapter The Impact of
  Modelling and Abstraction Concerns on Modern Programming Languages.
\newblock In {\em Topics in Information Systems\/} \cite{brodie1984conceptual},
  1984.

\bibitem{pstar12}
Andre Luckow, Mark Santcroos, Andre Merzky, Ole Weidner, Pradeep Mantha, and
  Shantenu Jha.
\newblock P*: A model of pilot-abstractions.
\newblock {\em IEEE 8th International Conference on e-Science}, pages 1--10,
  2012.
\newblock \newline \url{http://dx.doi.org/10.1109/eScience.2012.6404423}.

\bibitem{Iivari2007}
Juhani Iivari.
\newblock A paradigmatic analysis of information systems as a design science.
\newblock {\em Scandinavian Journal of Information Systems}, 19:39--, 01 2007.
\newblock \url{https://aisel.aisnet.org/sjis/vol19/iss2/5/}.

\bibitem{dsr}
Alan~R. Hevner, Salvatore~T. March, Jinsoo Park, and Sudha Ram.
\newblock Design science in information systems research.
\newblock {\em MIS Quarterly}, 28(1):75--105, 2004.

\bibitem{10.2753/MIS0742-1222240302}
Ken Peffers, Tuure Tuunanen, Marcus Rothenberger, and Samir Chatterjee.
\newblock A design science research methodology for information systems
  research.
\newblock {\em J. Manage. Inf. Syst.}, 24(3):45–77, December 2007.

\bibitem{10.1007/978-3-642-29863-9_28}
Christian Sonnenberg and Jan vom Brocke.
\newblock Evaluations in the science of the artificial -- reconsidering the
  build-evaluate pattern in design science research.
\newblock In Ken Peffers, Marcus Rothenberger, and Bill Kuechler, editors, {\em
  Design Science Research in Information Systems. Advances in Theory and
  Practice}, pages 381--397, Berlin, Heidelberg, 2012. Springer Berlin
  Heidelberg.

\bibitem{oro17673}
Judith Segal.
\newblock Models of scientific software development.
\newblock In {\em SECSE 08, First International Workshop on Software
  Engineering in Computational Science and Engineering}, May 2008.
\newblock Workshop co-located with ICSE 08 http://icse08.upb.de/.

\bibitem{doi:10.1002/cpe.2897}
Shantenu Jha, Murray Cole, Daniel~S. Katz, Manish Parashar, Omer Rana, and Jon
  Weissman.
\newblock Distributed computing practice for large-scale science and
  engineering applications.
\newblock {\em Concurrency and Computation: Practice and Experience},
  25(11):1559--1585.

\bibitem{Mattson:2004:PPP:1406956}
Timothy Mattson, Beverly Sanders, and Berna Massingill.
\newblock {\em Patterns for Parallel Programming}.
\newblock Addison-Wesley Professional, first edition, 2004.

\bibitem{10.1145/3274770}
Micah Beck.
\newblock On the hourglass model.
\newblock {\em Commun. ACM}, 62(7):48–57, June 2019.

\bibitem{10.1145/1176617.1176622}
Joshua Bloch.
\newblock How to design a good api and why it matters.
\newblock In {\em Companion to the 21st ACM SIGPLAN Symposium on
  Object-Oriented Programming Systems, Languages, and Applications}, OOPSLA
  ’06, page 506–507, New York, NY, USA, 2006. ACM.

\bibitem{mylopoulos1992conceptual}
John Mylopoulos.
\newblock Conceptual modelling and telos.
\newblock \url{http://www.cs.toronto.edu/~jm/2507S/Readings/CM+Telos.pdf},
  1992.

\bibitem{Johnson:2002:CMB:503355.503366}
Jeff Johnson and Austin Henderson.
\newblock Conceptual models: Begin by designing what to design.
\newblock {\em Interactions}, 9(1):25--32, January 2002.

\bibitem{brodie1984conceptual}
Michael~L. Brodie, John Mylopoulos, and Joachim~W. Schmidt.
\newblock {\em On Conceptual Modelling: Perspectives from Artificial
  Intelligence, Databases, and Programming Languages}.
\newblock Topics in Information Systems. Springer New York, 1984.

\bibitem{Bruegge:2009:OSE:1795808}
Bernd Bruegge and Allen~H. Dutoit.
\newblock {\em Object-Oriented Software Engineering Using UML, Patterns, and
  Java}.
\newblock Prentice Hall Press, Upper Saddle River, NJ, USA, 3rd edition, 2009.

\bibitem{919147}
M.~{Shaw}.
\newblock The coming-of-age of software architecture research.
\newblock In {\em Proceedings of the 23rd International Conference on Software
  Engineering. ICSE 2001}, pages 657--664a, May 2001.

\bibitem{alexander1977pattern}
C.~Alexander, P.D.A.C. Alexander, S.~Ishikawa, M.~Silverstein, M.~Jacobson,
  Center for Environmental~Structure, I.~Fiksdahl-King, and A.~Shlomo.
\newblock {\em A Pattern Language: Towns, Buildings, Construction}.
\newblock Center for Environmental Structure Berkeley, Calif: Center for
  Environmental Structure series. OUP USA, 1977.

\bibitem{Beck1987}
Kent Beck and Ward Cunningham.
\newblock Using pattern languages for object oriented programs.
\newblock In {\em Conference on Object-Oriented Programming, Systems,
  Languages, and Applications (OOPSLA)}, 1987.

\bibitem{Parnas:1972:CUD:361598.361623}
David~Lorge Parnas.
\newblock On the criteria to be used in decomposing systems into modules.
\newblock {\em Commun. ACM}, 15(12):1053--1058, December 1972.

\bibitem{Parnas1971}
David~Lorge Parnas.
\newblock {Information distribution aspects of design methodology}.
\newblock {\em Methods}, 4(5):6--7, 1971.

\bibitem{Garlan:1994:ISA:865128}
David Garlan and Mary Shaw.
\newblock An introduction to software architecture.
\newblock Technical report, Pittsburgh, PA, USA, 1994.

\bibitem{foster2003grid}
I.~Foster and C.~Kesselman.
\newblock {\em The Grid 2: Blueprint for a New Computing Infrastructure}.
\newblock ISSN. Elsevier Science, 2003.

\bibitem{toward-hpc-abds}
Judy Qiu, Shantenu Jha, Andre Luckow, and Geoffrey~C. Fox.
\newblock Towards hpc-abds: An initial high-performance big data stack.
\newblock In {\em Proceedings of ACM Big Data Interoperability Framework
  Workshop}, 2015.

\bibitem{10.2307/258557}
Kathleen~M. Eisenhardt.
\newblock Building theories from case study research.
\newblock {\em The Academy of Management Review}, 14(4):532--550, 1989.

\bibitem{Jain91}
Raj Jain.
\newblock {\em The art of computer systems performance analysis - techniques
  for experimental design, measurement, simulation, and modeling.}
\newblock Wiley professional computing. Wiley, 1991.

\bibitem{ferrari1978computer}
D.~Ferrari.
\newblock {\em Computer Systems Performance Evaluation}.
\newblock Prentice-Hall, 1978.

\bibitem{Gray:1992:BHD:530588}
Jim Gray.
\newblock {\em Benchmark Handbook: For Database and Transaction Processing
  Systems}.
\newblock Morgan Kaufmann, San Francisco, CA, USA, 1992.

\bibitem{DBLP:journals/corr/abs-1801-08648}
Andr{\'{e}} Luckow, George Chantzialexiou, and Shantenu Jha.
\newblock Pilot-streaming: {A} stream processing framework for high-performance
  computing.
\newblock {\em IEEE eScience International Conference}, abs/1801.08648, 2018.

\bibitem{5388337}
W.~{Buchholz}.
\newblock A synthetic job for measuring system performance.
\newblock {\em IBM Systems Journal}, 8(4):309--318, 1969.

\bibitem{7004228}
J.~W. {Anderson}, K.~E. {Kennedy}, L.~B. {Ngo}, A.~{Luckow}, and A.~W. {Apon}.
\newblock Synthetic data generation for the internet of things.
\newblock In {\em 2014 IEEE International Conference on Big Data (Big Data)},
  2014.

\bibitem{MERZKY2018329}
Andre Merzky, Ming~Tai Ha, Matteo Turilli, and Shantenu Jha.
\newblock Synapse: Synthetic application profiler and emulator.
\newblock {\em Journal of Computational Science}, 27:329 -- 344, 2018.

\bibitem{hoisie18}
Adolfy Hoisie.
\newblock Performance modeling overview.
\newblock Talk at PAM 2018: Performance Analysis and Modeling Workshop:
  \url{https://indico.bnl.gov/event/3950/contributions/12021/attachments/10817/13215/Talk_at_the_Perf_Workshop_Feb_2018.pdf
  }, 2018.

\bibitem{bordgida1984}
A.~Bordgida, J.~Mylopoulos, and H.~K.~T. Wong.
\newblock {\em On Conceptual Modelling: Perspectives from Artificial
  Intelligence, Databases, and Programming Languages}, chapter
  Generalization/Specialization as a Basis for Software Specification.
\newblock In {\em Topics in Information Systems\/} \cite{brodie1984conceptual},
  1984.

\bibitem{stat_model}
Danilo Bzdok, Naomi Altman, and Martin Krzywinski.
\newblock Statistics versus machine learning.
\newblock {\em Nature Methods}, 15:233 EP --, 04 2018.

\bibitem{Selinger:1979:APS:582095.582099}
P.~Griffiths Selinger, M.~M. Astrahan, D.~D. Chamberlin, R.~A. Lorie, and T.~G.
  Price.
\newblock Access path selection in a relational database management system.
\newblock In {\em Proceedings of the ACM SIGMOD International Conference on
  Management of Data}, SIGMOD '79, NY, NY, USA, 1979. ACM.

\bibitem{Amdahl:1967:VSP:1465482.1465560}
Gene~M. Amdahl.
\newblock Validity of the single processor approach to achieving large scale
  computing capabilities.
\newblock In {\em Proceedings of the April 18-20, 1967, Spring Joint Computer
  Conference}, AFIPS '67 (Spring), pages 483--485, New York, NY, USA, 1967.
  ACM.

\bibitem{DBLP:journals/corr/abs-1902-10810}
Geoffrey~C. Fox, James~A. Glazier, J.~C.~S. Kadupitiya, Vikram Jadhao, Minje
  Kim, Judy Qiu, James~P. Sluka, Endre~T. Somogyi, Madhav Marathe, Abhijin
  Adiga, Jiangzhuo Chen, Oliver Beckstein, and Shantenu Jha.
\newblock Learning everywhere: Pervasive machine learning for effective
  high-performance computation.
\newblock {\em CoRR}, abs/1902.10810, 2019.

\bibitem{NAP21886}
Engineering National Academies~of Sciences and Medicine.
\newblock {\em Future Directions for NSF Advanced Computing Infrastructure to
  Support U.S. Science and Engineering in 2017-2020}.
\newblock The National Academies Press, Washington, DC, 2016.

\bibitem{dwarfs}
Krste Asanović, Ras Bodik, Bryan~Christopher Catanzaro, Joseph~James Gebis,
  Parry Husbands, Kurt Keutzer, David~A. Patterson, William~Lester Plishker,
  John Shalf, Samuel~Webb Williams, and Katherine~A. Yelick.
\newblock The landscape of parallel computing research: A view from berkeley.
\newblock Technical Report UCB/EECS-2006-183, EECS Department, University of
  California, Berkeley, Dec 2006.

\bibitem{doi:10.1002/cpe.4032}
Shantenu Jha, Daniel~S. Katz, Andre Luckow, Neil Chue~Hong , Omer Rana, and
  Yogesh Simmhan.
\newblock Introducing distributed dynamic data-intensive (d3) science:
  Understanding applications and infrastructure.
\newblock {\em Concurrency and Computation: Practice and Experience}, 29(8),
  2017.

\bibitem{bigdata-ogres}
Geoffrey~C. Fox, Shantenu Jha, Judy Qiu, and Andre Luckow.
\newblock Towards an understanding of facets and exemplars of big data
  applications.
\newblock In {\em Proceedings of Beowulf'14}, Annapolis, MD, USA, 2014. ACM.

\bibitem{fox_bigdata_benchmarks}
Geoffrey~C. Fox, Shantenu Jha, Judy Qiu, and Andre Luckow.
\newblock A systematic approach to big data benchmarks.
\newblock In Lucio Grandinetti, Gerhard Joubert, Marcel Kunze, and Valerio
  Pascucci, editors, {\em Big Data and High Performance Computing}, volume~24,
  pages 47--66. IOS Press, München, 2015.
\newblock \url{http://dx.doi.org/10.3233/978-1-61499-583-8-47}.

\bibitem{nist-uc}
NIST BigData~Working Group.
\newblock \url{http://bigdatawg.nist.gov/usecases.php}, 2019.

\bibitem{repex_ptrsa}
Andre Luckow, Shantenu Jha, Joohyun Kim, Andre Merzky, and Bettina Schnor.
\newblock {Adaptive Replica-Exchange Simulations}.
\newblock {\em Royal Society Philosophical Transactions A}, pages 2595--2606,
  jun 2009.

\bibitem{doi:10.1021/ct4007037}
David~W. Wright, Benjamin~A. Hall, Owain~A. Kenway, Shantenu Jha, and Peter~V.
  Coveney.
\newblock Computing clinically relevant binding free energies of hiv-1 protease
  inhibitors.
\newblock {\em Journal of Chemical Theory and Computation}, 10(3):1228--1241,
  2014.
\newblock PMID: 24683369.

\bibitem{El-Khamra:2009:DAD:1555301.1555304}
Yaakoub El-Khamra and Shantenu Jha.
\newblock Developing autonomic distributed scientific applications: A case
  study from history matching using ensemble kalman-filters.
\newblock In {\em Proceedings of the 6th International Conference Industry
  Session on Grids Meets Autonomic Computing}, GMAC '09, pages 19--28, New
  York, NY, USA, 2009. ACM.

\bibitem{dare}
Sharath Maddineni, Joohyun Kim, Yaakoub El-Khamra, and Shantenu Jha.
\newblock Distributed application runtime environment (dare): A standards-based
  middleware framework for science-gateways.
\newblock {\em Journal of Grid Computing}, 10(4):647--664, 2012.

\bibitem{hpc-abds}
G.~C. {Fox}, J.~{Qiu}, S.~{Kamburugamuve}, S.~{Jha}, and A.~{Luckow}.
\newblock Hpc-abds high performance computing enhanced apache big data stack.
\newblock In {\em 2015 15th IEEE/ACM International Symposium on Cluster, Cloud
  and Grid Computing}, pages 1057--1066, May 2015.

\bibitem{Paraskevakos:2018:TAM:3225058.3225128}
Ioannis Paraskevakos, Andre Luckow, Mahzad Khoshlessan, George Chantzialexiou,
  Thomas~E. Cheatham, Oliver Beckstein, Geoffrey~C. Fox, and Shantenu Jha.
\newblock Task-parallel analysis of molecular dynamics trajectories.
\newblock In {\em Proceedings of the 47th International Conference on Parallel
  Processing}, ICPP 2018, New York, NY, USA, 2018. ACM.

\bibitem{Mantha:2012:PEF:2287016.2287020}
Pradeep~Kumar Mantha, Andre Luckow, and Shantenu Jha.
\newblock {Pilot-MapReduce: An Extensible and Flexible MapReduce Implementation
  for Distributed Data}.
\newblock In {\em {Proceedings of third international workshop on MapReduce and
  its Applications}}, MapReduce '12, pages 17--24, New York, NY, USA, 2012.
  ACM.

\bibitem{DBLP:journals/corr/JhaQLMF14}
Shantenu Jha, Judy Qiu, Andr{\'e} Luckow, Pradeep~Kumar Mantha, and
  Geoffrey~Charles Fox.
\newblock A tale of two data-intensive paradigms: Applications, abstractions,
  and architectures.
\newblock {\em Proceedings of 3rd IEEE Internation Congress of Big Data},
  abs/1403.1528, 2014.

\bibitem{Dean:2008:MSD:1327452.1327492}
Jeffrey Dean and Sanjay Ghemawat.
\newblock Mapreduce: simplified data processing on large clusters.
\newblock {\em Commun. ACM}, 51(1):107--113, January 2008.

\bibitem{sutherland1966line}
William~Robert Sutherland.
\newblock {\em The on-line graphical specification of computer procedures.}
\newblock PhD thesis, MIT, 1966.

\bibitem{301911}
D.~C. {DiNucci} and R.~G. {Babb}.
\newblock Design and implementation of parallel programs with lgdf2.
\newblock In {\em Digest of Papers. COMPCON Spring 89. Thirty-Fourth IEEE
  Computer Society International Conference: Intellectual Leverage}, pages
  102--107, Feb 1989.

\bibitem{dryad}
Michael Isard, Mihai Budiu, Yuan Yu, Andrew Birrell, and Dennis Fetterly.
\newblock {Dryad: distributed data-parallel programs from sequential building
  blocks}.
\newblock {\em SIGOPS Oper. Syst. Rev.}, 41(3):59--72, 2007.

\bibitem{Ekanayake:2010:TRI:1851476.1851593}
Jaliya Ekanayake, Hui Li, Bingjing Zhang, Thilina Gunarathne, Seung-Hee Bae,
  Judy Qiu, and Geoffrey Fox.
\newblock Twister: A runtime for iterative mapreduce.
\newblock In {\em Proceedings of the 19th ACM International Symposium on High
  Performance Distributed Computing}, HPDC '10, pages 810--818, New York, NY,
  USA, 2010. ACM.

\bibitem{streaming2015}
Geoffrey Fox, Shantenu Jha, and Lavanya Ramakrishnan.
\newblock Stream 2015 final report.
\newblock \url{http://streamingsystems.org/finalreport.pdf}, 2015.

\bibitem{prio-dnn}
Anand Jayarajan, Jinliang Wei, Garth Gibson, Alexandra Fedorova, and Gennady
  Pekhimenko.
\newblock Priority-based parameter propagation for distributed dnn training.
\newblock In {\em Proceedings of SysML}, 05 2019.

\bibitem{saga_bigjob_condor_cloud}
Andre Luckow, Lukas Lacinski, and Shantenu Jha.
\newblock {SAGA BigJob: An Extensible and Interoperable Pilot-Job Abstraction
  for Distributed Applications and Systems}.
\newblock In {\em {The 10th IEEE/ACM International Symposium on Cluster, Cloud
  and Grid Computing}}, pages 135--144, 2010.

\bibitem{glidein}
James Frey, Todd Tannenbaum, Miron Livny, Ian Foster, and Steven Tuecke.
\newblock Condor-g: A computation management agent for multi-institutional
  grids.
\newblock {\em Cluster Computing}, 5(3):237--246, July 2002.

\bibitem{turilli2017comprehensive}
Matteo Turilli, Mark Santcroos, and Shantenu Jha.
\newblock A comprehensive perspective on pilot-job systems.
\newblock {\em ACM Comput. Surv.}, 51(2):43:1--43:32, April 2018.

\bibitem{pilotdata}
Andre Luckow, Mark Santcroos, Ashley Zebrowski, and Shantenu Jha.
\newblock Pilot-data: An abstraction for distributed data.
\newblock {\em Journal of Parallel and Distributed Computing}, 2014.

\bibitem{DBLP:journals/corr/LuckowMJ15}
Andre Luckow, Pradeep~Kumar Mantha, and Shantenu Jha.
\newblock Pilot-abstraction: {A} valid abstraction for data-intensive
  applications on hpc, hadoop and cloud infrastructures?
\newblock {\em CoRR}, abs/1501.05041, 2015.

\bibitem{2016arXiv160200345L}
A.~{Luckow}, I.~{Paraskevakos}, G.~{Chantzialexiou}, and S.~{Jha}.
\newblock {Hadoop on HPC: Integrating Hadoop and Pilot-based Dynamic Resource
  Management}.
\newblock {\em IEEE International Workshop on High-Performance Big Data
  Computing in conjunction with The 30th IEEE International Parallel and
  Distributed Processing Symposium (IPDPS 2016)}, 2016.

\bibitem{gamma1994design}
Erich Gamma, Richard Helm, Ralph Johnson, and John~M. Vlissides.
\newblock {\em Design Patterns: Elements of Reusable Object-Oriented Software}.
\newblock Addison-Wesley Professional, 1 edition, 1994.

\bibitem{saga-x}
Andre Merzky, Ole Weidner, and Shantenu Jha.
\newblock {SAGA}: A standardized access layer to heterogeneous distributed
  computing infrastructure\.
\newblock {\em Software-X}, 2015.
\newblock DOI: 10.1016/j.softx.2015.03.001.

\bibitem{uml}
{Object Management Group (OMG)}.
\newblock Unified modeling language specification version 2.5.1.
\newblock \url{https://www.omg.org/spec/UML/2.5.1/}, 2017.

\bibitem{async_repex11}
Abhinav Thota, Andre Luckow, and Shantenu Jha.
\newblock {Efficient large-scale Replica-Exchange Simulations on Production
  Infrastructure}.
\newblock {\em {Philosophical Transactions of the Royal Society A:
  Mathematical, Physical and Engineering Sciences}}, 369(1949):3318--3335,
  2011.

\bibitem{luckow2019performance}
Andre Luckow and Shantenu Jha.
\newblock Performance characterization and modeling of serverless and hpc
  streaming applications.
\newblock In {\em Proceedings of StreamML Workshop at IEEE International
  Conference on Big Data (IEEE BigData 2019)}, 2019.

\bibitem{10.1007/978-3-030-10632-4_4}
Andre Merzky, Matteo Turilli, Manuel Maldonado, Mark Santcroos, and Shantenu
  Jha.
\newblock Using pilot systems to execute many task workloads on supercomputers.
\newblock In {\em Job Scheduling Strategies for Parallel Processing}, pages
  61--82, Cham, 2019. Springer International Publishing.

\bibitem{hauck2014automated}
M.~Hauck.
\newblock {\em Automated Experiments for Deriving Performance-relevant
  Properties of Software Execution Environments:}.
\newblock The Karlsruhe Series on Software Design and Quality. KIT Scientific
  Publishing, 2014.

\bibitem{diss_mods_00016245}
Jan Waller.
\newblock Performance benchmarking of application monitoring frameworks.
\newblock \url{https://macau.uni-kiel.de/receive/diss_mods_00016245}, 2015.

\bibitem{10.1145/2484762.2484819}
Jack~A. Smith, Melissa Romanus, Pradeep~Kumar Mantha, Yaakoub El~Khamra,
  Thomas~C. Bishop, and Shantenu Jha.
\newblock Scalable online comparative genomics of mononucleosomes: A bigjob.
\newblock In {\em Proceedings of the Conference on Extreme Science and
  Engineering Discovery Environment: Gateway to Discovery}, XSEDE ’13, NY,
  NY, USA, 2013. ACM.

\bibitem{turilli2019middleware}
Matteo Turilli, Vivek Balasubramanian, Andre Merzky, Ioannis Paraskevakos, and
  Shantenu Jha.
\newblock Middleware building blocks for workflow systems.
\newblock {\em Computing in Science \& Engineering (CiSE) special issue on
  Incorporating Scientific Workflows in Computing Research Processes}.
\newblock 10.1109/MCSE.2019.2920048 (2019).

\bibitem{cloud_book}
Shantenu Jha, {Daniel S.} Katz, Andre Luckow, Andre Merzky, and Katerina
  Stamou.
\newblock {\em Understanding Scientific Applications for Cloud Environments},
  pages 345--371.
\newblock John Wiley and Sons, 1 2011.

\bibitem{leaky}
Joel Spolsky.
\newblock The law of leaky abstractions.
\newblock {\em Joel on Software: And on Diverse and Occasionally Related
  Matters}, 01 2002.

\end{thebibliography}

\end{document}